\documentclass{emulateapj}

\shorttitle{Progating Wave Phenomena in the Lower Solar Atmosphere}
\shortauthors{D.B. Jess et al.}

\begin{document}


\title{Propagating Wave Phenomena Detected in Observations and Simulations 
of the Lower Solar Atmosphere}


\author{D. B. Jess, S. Shelyag, M. Mathioudakis, P. H. Keys}
\affil{Astrophysics Research Centre, School of Mathematics and Physics, Queen's University Belfast, 
Belfast, BT7~1NN, Northern Ireland, U.K.}

\email{d.jess@qub.ac.uk}

\author{D. J. Christian}
\affil{Department of Physics and Astronomy, California State University Northridge, Northridge, 
CA 91330, U.S.A.}

\and

\author{F. P. Keenan}
\affil{Astrophysics Research Centre, School of Mathematics and Physics, Queen's University Belfast, 
Belfast, BT7~1NN, Northern Ireland, U.K.}


%\altaffiltext{1}{Visiting Astronomer, Cerro Tololo Inter-American Observatory.
%CTIO is operated by AURA, Inc.\ under contract to the National Science
%\altaffiltext{2}{Society of Fellows, Harvard University.}
%\altaffiltext{3}{present address: Center for Astrophysics,
%   60 Garden Street, Cambridge, MA 02138}
%\altaffiltext{4}{Visiting Programmer, Space Telescope Science Institute}
%\altaffiltext{5}{Patron, Alonso's Bar and Grill}


\begin{abstract}
We present high-cadence observations and simulations of the solar 
photosphere, obtained using the Rapid Oscillations in the Solar 
Atmosphere imaging system and the MuRAM magneto-hydrodynamic 
code, respectively. Each dataset demonstrates a wealth of magneto-acoustic 
oscillatory behaviour, visible as periodic intensity fluctuations with periods 
in the range 110 -- 600~s. Almost no propagating waves with periods less 
than 140~s and 110~s are detected in the observational and simulated 
datasets, respectively. High concentrations of power are found in 
highly magnetised regions, such as magnetic bright points and intergranular 
lanes. Radiative diagnostics of the photospheric simulations replicate
our observational results, confirming that the current breed of 
magneto-hydrodynamic simulations are able to accurately represent the 
lower solar atmosphere. All observed oscillations are generated as a 
result of naturally occurring magnetoconvective processes, with no specific input 
driver present. Using contribution functions extracted from our numerical 
simulations, we estimate minimum G-band and 4170{\,}{\AA} continuum 
formation heights of 100~km and 25~km, respectively. Detected magneto-acoustic 
oscillations exhibit a dominant phase delay of $-8{\degr}$ between the G-band and 
4170{\,}{\AA} continuum observations, suggesting the presence of 
upwardly propagating waves.  More than 73\% of MBPs (73\% from observations, 
96\% from simulations) display upwardly propagating wave phenomena, 
suggesting the abundant nature of oscillatory behaviour detected higher 
in the solar atmosphere may be traced back to magnetoconvective processes 
occurring in the upper layers of the Sun's convection zone. 
\end{abstract}





\section{Introduction}

\begin{figure*}
\epsscale{1.0}
\plotone{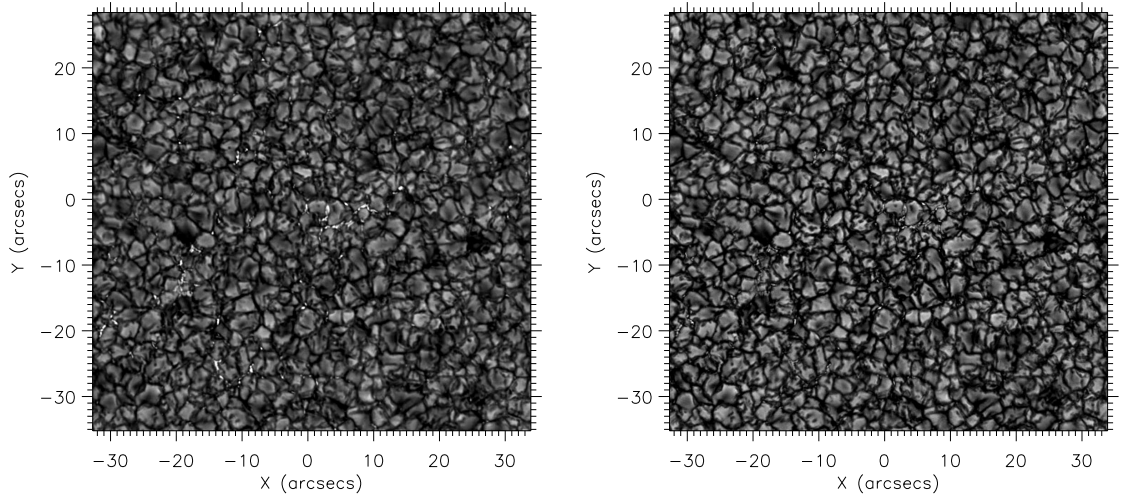}
\caption{Simultaneous ROSA images acquired through G-band (left) and 4170{\,}{\AA} continuum 
(right) filters at 14:06:21~UT on 2009 May 28. Axes are in heliocentric arcseconds. Note 
the abundance of MBPs, particularly visible as intensity enhancements in the G-band.
\label{fig1}}
\end{figure*}

Following the discovery of solar oscillations in the 1960s \citep[]{Lei60}, there has been a 
multitude of observational evidence brought forward verifying the existence of wave motion in the 
solar atmosphere \citep[]{Ste74}. In more recent years, the role of these oscillations is often 
closely linked to the search for efficient heating mechanisms of the upper solar atmosphere 
\citep[see e.g.][]{Str08, Cau09, Tar09}. 

Wave phenomena in the lower solar atmosphere (photosphere and chromosphere) can exist in 
a variety of different structures, and importantly exhibit various modes of oscillation \citep[]{Edw83}. 
Magneto-acoustic, Alfv{\'{e}}n, sausage, and kink modes have recently been observed in 
a wide range of solar features \citep[]{DeP07, Jes07, Jes09, Mor11}. Importantly, however, is 
the fact that the majority of these wave phenomena exist in structures which are highly magnetic 
in nature, such as active regions, pores, and magnetic bright points (MBPs). Indeed, 
\citet{Rim95} found photospheric oscillations to preferentially exist in intrinsically magnetic 
intergranular lanes. In the Sun's 
atmosphere, magnetic field lines clump into tight bundles forming highly-structured wave 
guides, which allow oscillatory motion to be readily channelled along these structures 
\citep[]{She10, Fed11}. At any one time it is estimated that up to 2.2{\%} of the entire solar surface is 
covered by MBPs \citep[]{San10}, and with their intrinsic magnetic field strengths exceeding one 
kiloGauss \citep[]{Jes10a}, they may provide the basis of a suitable mechanism for efficient 
energy transport into the higher layers of the Sun's atmosphere.

Acoustic events occurring in the lower solar atmosphere are often closely examined in 
an attempt to determine the mechanisms behind, and the energy fluxes associate with, 
such wave phenomena. \citet{Hoe02} utilised high-resolution ground-based observations 
to investigate the relation between upwardly propagating acoustic waves and co-spatial 
chromospheric brightenings, while \citet{Ste91} suggested how acoustic modes may be 
excited stochastically by non-adiabatic fluctuations in the gas pressure near the solar surface. 
Highly energetic acoustic waves have been detected by \citet{Bel10b} using 
spectropolarimetric data taken with the IMaX/SUNRISE balloon experiment, with energy 
fluxes exceeding 7000~W{\,}m$^{-2}$ in the low photosphere. 
Contrarily, \citet{Fos05} and \citet{Car07} examined the presence of acoustic oscillations in the solar 
chromosphere, and concluded that the energy flux of these waves was not 
sufficiently high to constitute the dominant atmospheric heating mechanism. However, due to 
telemetry restrictions, the use of data obtained from space-based instrumentation limited 
their search to oscillations with periods longer than 25~s. This may have resulted in a considerable 
fraction of energy, at short periods ($\sim$10~s), being overlooked \citep[]{DeF04, Has08}. 
Moreover, the contribution functions associated with the filter bandpasses used by 
\citet{Fos05} and \citet{Car07} are very 
extended, and must therefore be taken into further consideration \citep[]{Bel09}. 
Thus, the direct role of acoustic waves in the heating of the solar atmosphere remains 
uncertain \citep[]{Kal07, Bec08, Bel10a}. To understand the presence of wave 
phenomena in the Sun's outer atmosphere, it is imperative to study the photospheric 
counterpart of these oscillations, and establish how ubiquitous these fundamental 
waves are in the lower solar atmosphere. In this paper, we utilise a high cadence 
multi-wavelength data set to investigate propagating wave phenomena detected in observations 
and simulations of the lower solar atmosphere.

\section{Data Sets}
\subsection{Observations}
The observational data presented here are part of a sequence obtained during 
13:46 -- 14:47~UT on 2009 May 28, with the Dunn Solar Telescope at Sacramento 
Peak, New Mexico. We employ the Rapid Oscillations in the Solar Atmosphere 
\citep[ROSA;][]{Jes10b} six-camera system 
to image a $69.3''\times69.1''$ region positioned at solar disk center.   
A spatial sampling of $0.069''$ per pixel was used for the ROSA cameras, to 
match the telescope's diffraction-limited resolution in the blue continuum to that of 
the CCD. This results in images obtained at longer wavelengths being slightly 
oversampled. However, this was deemed desirable to keep the dimensions of the 
field-of-view the same for all ROSA cameras. 

During the observations, high-order adaptive optics \citep[]{Rim04} 
were used to correct wavefront deformations in real-time. The acquired images were 
further improved through use of the KISIP speckle reconstruction algorithm \citep[]{Wog08}, 
utilising $16 \rightarrow 1$ restorations. Combining consistently excellent atmospheric seeing 
conditions with speckle image reconstruction, a 2-pixel 
diffraction-limited resolution of $\approx$100~km is achieved for both G-band and 4170{\,}{\AA} 
blue continuum imaging. 
For the purposes of this paper, only observations acquired through the photospheric 
G-band and 4170{\,}{\AA} blue continuum filters will be presented, each with a 
post-reconstruction cadence of 0.528~s. The G-band and 4170{\,}{\AA} blue continuum 
filters have central wavelengths of 4305.5{\,}{\AA} and 4170.0{\,}{\AA}, in 
addition to full-width half-maximums of 9.2{\,}{\AA} and 52.0{\,}{\AA}, respectively. 
To insure accurate coalignment in all bandpasses, broadband time series were 
Fourier co-registered and de-stretched using a 
$40\times40$ grid, equating to a $\approx$1.7$''$ separation between spatial samples 
\citep[]{Jes07, Jes08}. 

\begin{figure*}
\epsscale{0.9}
\plotone{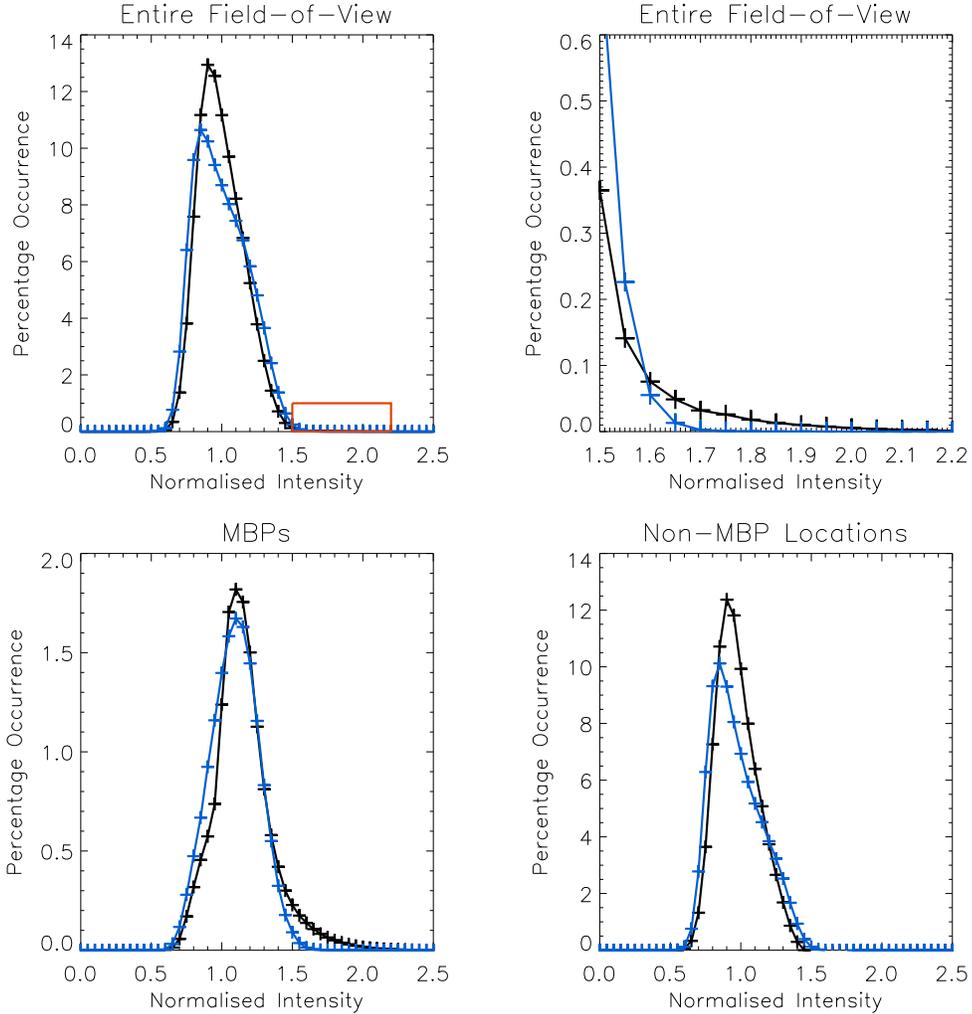}
\caption{{\it{Upper Left}}: Histograms of the number of pixels, as a percentage of the total number of pixels, 
versus their normalised intensity in the G-band (black) and 4170{\,}{\AA} continuum 
(blue) filtergrams, evaluated over the entire 61 minute duration of the data set 
($7.0{\times}10^{9}$ individual pixels). {\it{Upper Right}}: A zoom in of the intensity occurrences contained 
within the red box in the left panel. G-band intensities display an extended tail, well above the 
4170{\,}{\AA} continuum maximum, as a result of the presence of MBPs. {\it{Lower Left}}: Occurrence 
of intensities for isolated MBP structures, clearly displaying excess high-intensity measurements 
for G-band observations, also apparent in the upper-right panel. {\it{Lower Right}}: Intensity histograms 
for regions of the field-of-view not containing MBP structures. A narrowing of the 4170{\,}{\AA} continuum 
profile near its maximum occurrence, caused by the removal of MBP intensities, directly results in the 
reduction of its positive transmission skewness. 
\label{fig2}}
\end{figure*}

\subsection{Simulations}
We utilise the MuRAM \citep[]{Vog05} radiative magneto-hydrodynamic (MHD) code 
to investigate whether well-understood 
magnetoconvection processes can produce comparable phenomena to 
those detected in our observational time series. This code solves large-eddy radiative 
three-dimensional MHD equations on a Cartesian grid, and employs a fourth-order 
Runge-Kutta scheme to advance the numerical solution in time. The numerical domain 
has a physical size of $12\times12$~Mm$^{2}$ in the horizontal direction, 1.4~Mm in 
the vertical direction, and is resolved by $480\times480\times100$ grid cells, 
respectively. Our starting point for the simulations is a well-developed non-magnetic 
($B=0$) snapshot of photospheric convection taken approximately 2000~s 
(about 8 convective turnover timescales) from the initial plane-parallel model. A uniform 
vertical magnetic field of 200~G was introduced at this stage, and a sequence of 872 
snapshots recorded, each separated by a time interval of $\sim$2~s. 
The resulting sequence, which is used for further radiative diagnostics, 
covers approximately 30~min of physical time, corresponding to $\sim$4 -- 6 granular 
lifetimes. Full details of the simulation parameters and domain characteristics can be 
found in \citet{She11b}. 

\begin{figure*}
\epsscale{1.0}
\plotone{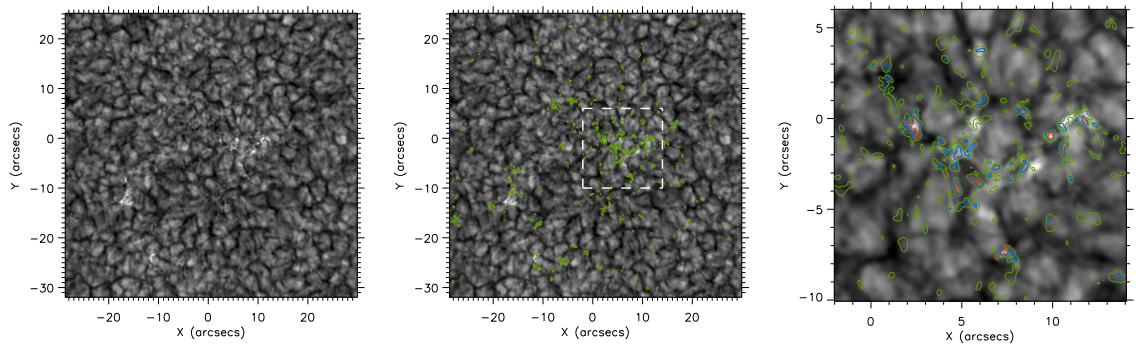}
\caption{A G-band intensity image, averaged over the entire 61~minute duration of the time series (left). 
Green contours (middle and right panels) highlight the locations where waves are observed to propagate 
(i.e. a non-zero phase shift) between the G-band and 4170{\,}{\AA} continuum bandpasses. These 
contours are generated irrespective of oscillation frequency, and so cannot be used to pinpoint where 
particular periodicities exist. Dashed white 
lines outline a sub-field, which is magnified in the right panel. Blue and red contours in the right 
panel represent propagating wave power which is four and six orders-of-magnitude, respectively, above the 
background. A preferential existence of propagating waves is found in locations of 
increased magnetic field strength, such as MBPs.
\label{fig3}}
\end{figure*}

The broadband filters used in the optical observations allow us to neglect the effects of 
line-of-sight velocities and magnetic fields in the radiative diagnostics. Due to a lack of 
detailed information on absorption line profiles contained within the filter bandpasses, 
precise computation of the wavelength dependence of emergent intensity within each 
filter is not necessary. Thus a simplified and computationally efficient approach of 
calculating the transmitted filter intensities is used. The methodology, similar to that 
used by \citet{afram1}, is based on the ATLAS9 spectral synthesis package of 
\citet{kurucz1}. Continuum and line opacity tables were calculated for a realistic solar chemical 
composition, incorporating a microturbulent velocity, $v_{\mathrm{turb}}=2$~km{\,}s$^{-1}$, 
and subsequently extracted from the ATLAS9 package using the subset codes 
XNFDF and DFSYNTHE. This procedure takes into account absorption 
by diatomic CH molecules. To avoid some of the problems highlighted 
by \citet{afram1}, the line opacity wavelength grid between 4290 -- 4400~{\AA} was made 
denser, and centered at the wavelength of the G-band filter transmission maximum. 
The DELO solver \citep{rees1} was used to numerically integrate the corresponding radiative 
transport equations for all $480 \times 480$ vertical rays of the 872 photospheric snapshots,
allowing the computation of the intensity spectra within the 4170{\,}{\AA}  and G-band filter 
bandpasses. Calculated 
spectra were then convolved with the G-band and 4170{\,}{\AA} filter transmission functions to
obtain a true intensity time series, which is directly comparable with our observational data. 
A more in-depth discussion of the methodology outlined here can be 
found in \citet{She11c}.

A natural feature of the numerical simulation is the ability to determine wavelength-dependent 
contribution functions 
of emergent radiation for the filter bandpasses which reside within its 1.4~Mm vertical 
domain. Using previously obtained continuum intensities, as well as continuum and line band 
absorption coefficients, we determined the corresponding continuum and ``band 
depression'' (defined as $\left(I-I_c\right)/I_c$, where $I$ and $I_c$ are the 
emergent and continuum intensities, respectively) contribution functions for our 
observational G-band (4305.5{\,}{\AA} central wavelength, 9.2{\,}{\AA} bandpass) 
and 4170{\,}{\AA} continuum (4170.0{\,}{\AA} central wavelength, 52.0{\,}{\AA} bandpass) 
filters. The band depression contribution functions were calculated 
from the Rosseland means of the continuum and line absorption coefficients, and based on
the definition provided by \citet{gurtovenko} and \citet{magain}.

%% In a manner similar to \objectname authors can provide links to dataset
%% hosted at participating data centers via the \dataset{} command.  The
%% second curly bracket argument is printed in the text while the first
%% parentheses argument serves as the valid data set identifier.  Large
%% lists of data set are best provided in a table (see Table 3 for an example).
%% Valid data set identifiers should be obtained from the data center that
%% is currently hosting the data.
%%
%% Note that AASTeX interprets everything between the curly braces in the 
%% macro as regular text, so any special characters, e.g. "#" or "_," must be 
%% preceded by a backslash. Otherwise, you will get a LaTeX error when you 
%% compile your manuscript.  Special characters do not 
%% need to be escaped in the optional, square-bracket argument.

%% In this section, we use  the \subsection command to set off
%% a subsection.  \footnote is used to insert a footnote to the text.


%% This section also includes several of the displayed math environments
%% mentioned in the Author Guide.

\section{Analysis and Discussion} 
\label{analysis}
Overall, the Sun was very quiet on 2009 May 28, with no active regions, pores 
or large-scale magnetic activity present on disk. However, our ROSA field-of-view, 
positioned at the center of the solar disk, contained numerous collections of MBPs, 
providing an ideal opportunity to examine these kiloGauss 
structures without the line-of-sight effects associated with off disc-center observations. 
Figure~\ref{fig1} shows simultaneous G-band and 4170{\,}{\AA} continuum snapshots, 
revealing a wealth of MBPs, which are particularly visible as intensity enhancements 
in the G-band filtergram.

\subsection{Filter Transmission Profiles}
Following speckle image reconstruction, and to assist identification of inherently 
bright and dim structures, each of the G-band and 4170{\,}{\AA} 
continuum images were normalised to their mean. To examine the 
throughput of each filter, an intensity histogram was created by binning 
pixel intensities into intervals of 0.05 normalised units. The upper panels of 
Figure~\ref{fig2} plot histograms of the number of pixels 
(as a percentage of the total number of pixels) versus normalised intensity, 
summed over the entire 
61 minute duration of the data set (in excess of $7.0 \times 10^{9}$ individual 
pixels). It is clear from the upper-left panel of Figure~\ref{fig2} that the G-band mean 
is close to its modal value. However, for the 4170{\,}{\AA} continuum 
observations, the modal value of the images lies below the filter average, suggesting 
a degree of positive skew in the transmitted intensities. A measurement of the 
skewness can be determined through the calculation of Fisher and Pearson 
coefficients of skewness, given by,
\begin{equation}
\mathrm{Fisher~Coefficient} = 
\frac{\frac{1}{N}\sum^{N}_{i=1}(x_{i} - \bar{x})^{3}}
{\left(\frac{1}{N}\sum^{N}_{i=1}(x_{i} - \bar{x})^{2}\right)^{3/2}} \ {\mathrm{,~and}} 
\end{equation}
\begin{equation}
\mathrm{Pearson~Coefficient} = 
\frac{3\left(\frac{1}{N}\sum^{N}_{i=1}(x_{i}) - \tilde{x}\right)}
{\left(\frac{1}{N}\sum^{N}_{i=1}(x_{i} - \bar{x})^{2}\right)^{1/2}} \ {\mathrm{,}}
\end{equation}
where $x_{i}$ is the pixel intensity, $\bar{x}$ is the mean, and $\tilde{x}$ is the median, 
calculated over $N$ independent measurements. While a symmetric distribution will 
yield a skewness coefficient equal to zero, Fisher and Pearson statistics 
for the 4170{\,}{\AA} continuum image sequence yield skewness coefficients of 
0.39 and 0.36, respectively. A potential weakness of the Fisher coefficient is its 
sensitivity to anomalous measurements occurring at the extreme ends of the 
intensity scale. However, consistency with the more robust Pearson coefficient 
indicates the presence of a well defined, and positively skewed, 4170{\,}{\AA} 
continuum transmission profile. 

\begin{figure*}
\epsscale{0.75}
\plotone{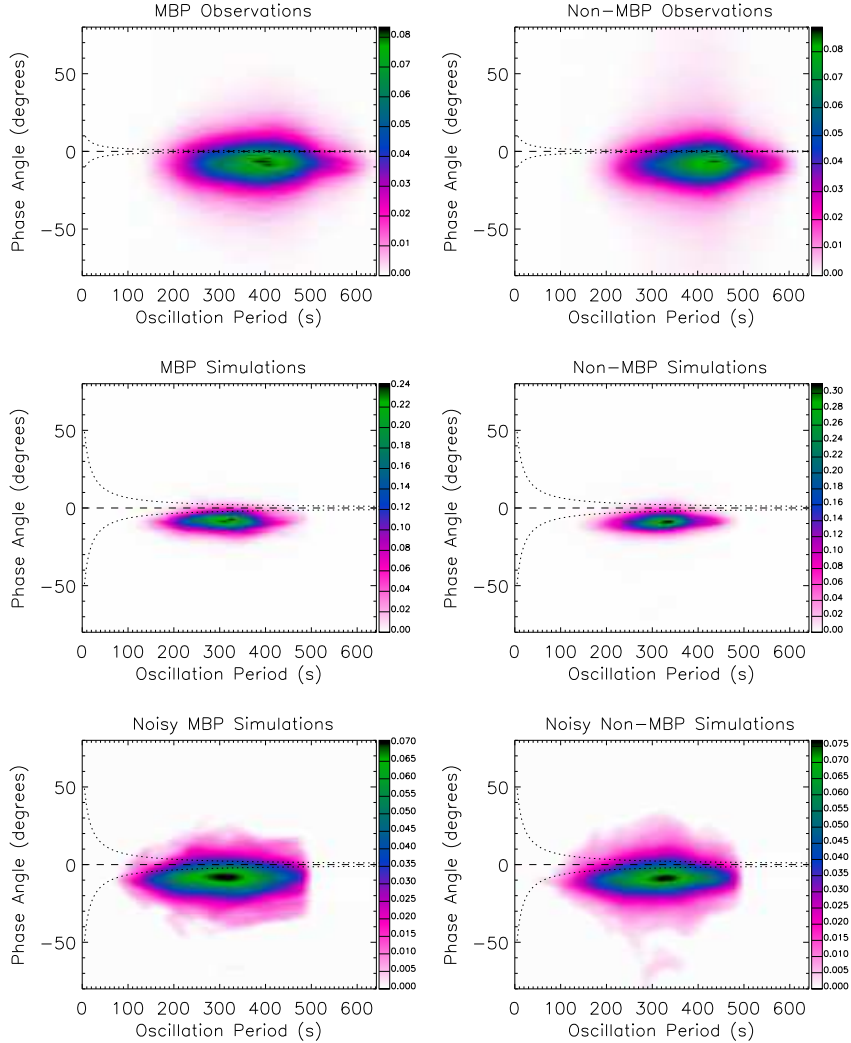}
\caption{Occurrence of oscillations simultaneously visible in both G-band and 4170{\,}{\AA} 
continuum images, as a function of oscillation period and phase angle for MBP (left), and 
non-MBP (right) regions. The upper panels 
represent oscillations occurring in the observational time series, while the middle and lower 
panels display oscillatory phenomena detected in the noise-free and 5\% added noise 
simulated datasets, respectively. The color scale 
represents the number of detections as a percentage of the total events 
with an associated coherence level exceeding 0.85. Consequently, the sum of each panel 
equals 100\%. Dotted lines highlight a region inside which detections become unreliable due 
to cadence restrictions (0.528~s and 2~s for observational and simulated time series, 
respectively), while a horizontal dashed line represents a phase angle of 0 degrees. A 
negative phase shift implies that an oscillation detected in the 4170{\,}{\AA} continuum 
leads one also detected in the G-band.
\label{fig4}}
\end{figure*}

Through examination of Figure~\ref{fig1}, it is clear that MBPs do 
not exhibit the same large contrast in 4170{\,}{\AA} continuum images that they 
do in simultaneous G-band snapshots. This can also be verified from the upper-right 
panel of Figure~\ref{fig2}, where the occurrence of 4170{\,}{\AA} continuum 
intensities falls to zero beyond 1.7 times the mean, compared with G-band 
intensities often exceeding twice the image mean. Interestingly, an 
intensity histogram of isolated MBP structures (lower-left panel of Fig.~\ref{fig2}) 
reveals that the most commonly occurring MBP intensity is identical 
($\approx$1.1 times the image mean) in both bandpasses. However, an extended 
high-intensity tail is pronounced in the G-band MBP histogram, resulting in a high 
feature contrast for these structures. This elevated contrast is due to its continuum 
formation level being depressed into deeper (and hotter) layers of the photosphere, 
where stronger magnetic fields and reduced CH abundances exist. 
An extended upper intensity range, and subsequently high feature 
contrast, of G-band MBP observations makes them ideal for the detection and 
tracking of such highly magnetic structures. As a result, we applied the MBP 
detection and tracking algorithms of \citet{Cro09, Cro10} and \citet{Keys11} 
to create binary maps 
for each of the G-band images. These binary images are used to pinpoint the 
exact locations within our field-of-view where MBPs exist, thus allowing their 
intensity fluctuations to be precisely examined as a function of time.

\begin{figure*}
\epsscale{1.0}
\plotone{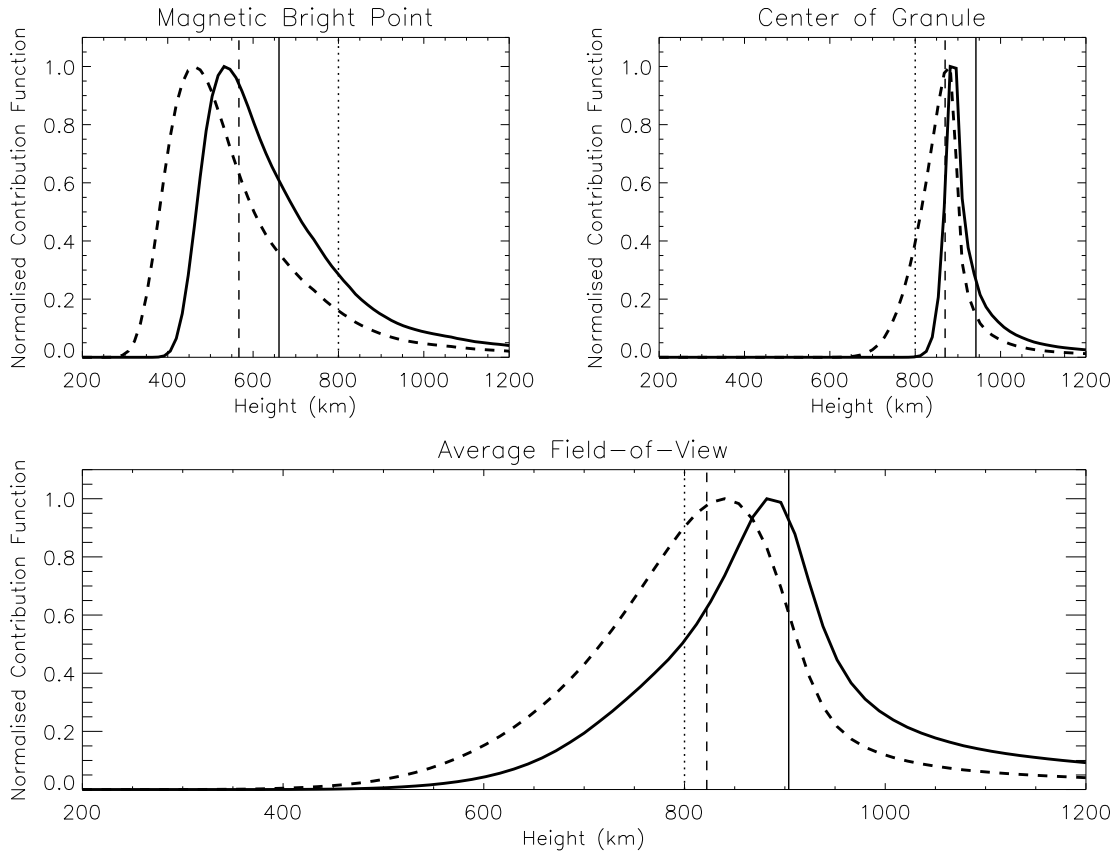}
\caption{The normalised filter contributions as a function of atmospheric height for 
G-band (solid line) and 4170{\,}{\AA} continuum (dashed line) bandpasses. The upper-left 
panel displays the contribution functions for a typical MBP, with the upper-right panel presenting 
contributions arising from the center of a granular cell. The bottom panel represents the 
contribution functions following a spatial averaging of the entire simulated field-of-view. 
A peak separation is clearly visible in the MBP case, while a peak overlap is present for the 
granular cell. However, due to an extended tail in all G-band contribution functions, 
a measurement of the center-of-gravity separation still yields a height difference 
in the case of a granular cell, even when the contribution peaks overlap. The atmospheric heights 
corresponding to the center-of-gravity of the G-band and 4170{\,}{\AA} continuum contribution 
functions, in addition to the location of the visible solar surface, are represented in each panel 
by solid, dashed and dotted vertical lines, respectively. A 
2-dimensional representation of these methods are shown in Figure~\ref{fig6}. 
\label{fig5}}
\end{figure*}

To investigate whether a shift of MBP intensities to lower values directly 
contributes to the positive skewness observed in the 4170{\,}{\AA} continuum 
transmission profile, the MBP binary maps were utilised to mask out locations 
containing these small-scale structures. The lower-right panel of Figure~\ref{fig2} 
displays the resulting intensity histogram for non-MBP areas. It is clear that the 
extended G-band high-intensity tail is no longer present, in addition to a substantially 
reduced profile width near the 4170{\,}{\AA} continuum maximum. Determination of 
the Fisher and Pearson statistics for the 4170{\,}{\AA} continuum 
image sequence containing no MBPs yield skewness coefficients of 0.20 and 
0.18, respectively. These values are significantly reduced from those derived 
from the entire field-of-view, indicating that the lower MBP intensities found in 
4170{\,}{\AA} continuum observations directly contributes to the positive skew 
found in the transmission profile.

\subsection{Propagating Wave Phenomena}
Following the methodology of \citet{Jes07, Jes07b}, the wavelet analysis routines 
of \citet{Tor98} were applied to both the G-band and 4170{\,}{\AA} continuum time series 
to search for the presence of oscillatory behaviour. Considerable 
oscillatory power, in both bandpasses, was present throughout the entire 
field-of-view, with enhanced concentrations at the locations of MBPs. Due to the 
image plate scales and camera dimensions being identical for G-band and 
4170{\,}{\AA} continuum data, no further 
co-alignment or de-stretching between data sets was required to provide accurate 
structural alignment. This can be verified visually through the examination 
of Figure~\ref{fig1}. Subsequently, to test the propagation of oscillatory phenomena 
between the two filters, coherence and phase information between simultaneous 
G-band and 4170{\,}{\AA} continuum pixels were extracted. The resulting data 
array contained oscillation period (1.056 \{Nyquist\} $\rightarrow$ 1830~seconds), 
power as a percentage above the background, coherence (0.0 for incoherent 
waves $\rightarrow$ 1.0 for coherent oscillations), and phase angle ($-180{\degr} 
\rightarrow +180{\degr}$) for each of the $7.0\times10^{9}$ imaging pixels. 
Following this nomenclature, a negative phase angle can be interpreted as an oscillation 
observed in the 4170{\,}{\AA} continuum leading one also detected in the G-band. 
A time-averaged G-band image is shown in the left panel of Figure~\ref{fig3}, 
contoured, in the middle panel, with the locations of propagating (i.e. a non-zero 
phase shift) waves. Approximately 1\% of the observational field-of-view contains 
propagating waves, with a preferential existence of these oscillations in locations of 
increased magnetic field strength, such as MBPs (right panel of Fig.~\ref{fig3}). 

To evaluate whether a preferential phase angle existed between G-band and 
4170{\,}{\AA} continuum oscillations, an occurrence plot, as a function of 
period and phase angle, was generated for all periodicities which displayed 
coherence values above 0.85. To differentiate between oscillations occurring 
in MBP structures from those occurring in the intergranular lanes, the MBP binary 
maps created above were utilised. By multiplying through each period, power, 
coherence, and phase angle value by the corresponding pixel binary, an 
occurrence plot based on where the oscillation was generated can be displayed. 
The upper panels of Figure~\ref{fig4} show the occurrence distributions 
for oscillations generated in both MBP and non-MBP regions. It is 
clear that a negative phase shift exists in both panels, with peak occurrences 
at a phase angle of $-8{\degr}$. Positive phase angles do exist, suggesting 
G-band oscillations lead their 4170{\,}{\AA} continuum counterparts in 
these instances. However, a clear majority of detections (73\% for MBP 
structures and 71\% for non-MBP regions) exhibit negative phase shifts, implying 
a preference for 4170{\,}{\AA} continuum oscillations to lead their G-band 
counterparts. A non-zero phase shift between the detected oscillations suggests that the observed 
wave phenomena are generated at distinctly different heights in the solar atmosphere, 
and implies the presence of propagating waves. Previous observational 
and theoretical studies of 
propagating wave phenomena have indicated their potential for 
transporting energy away from the solar surface, and into higher 
atmospheric layers \citep[e.g.][]{Car97, Vec07}.

\begin{figure*}
\epsscale{0.85}
\plotone{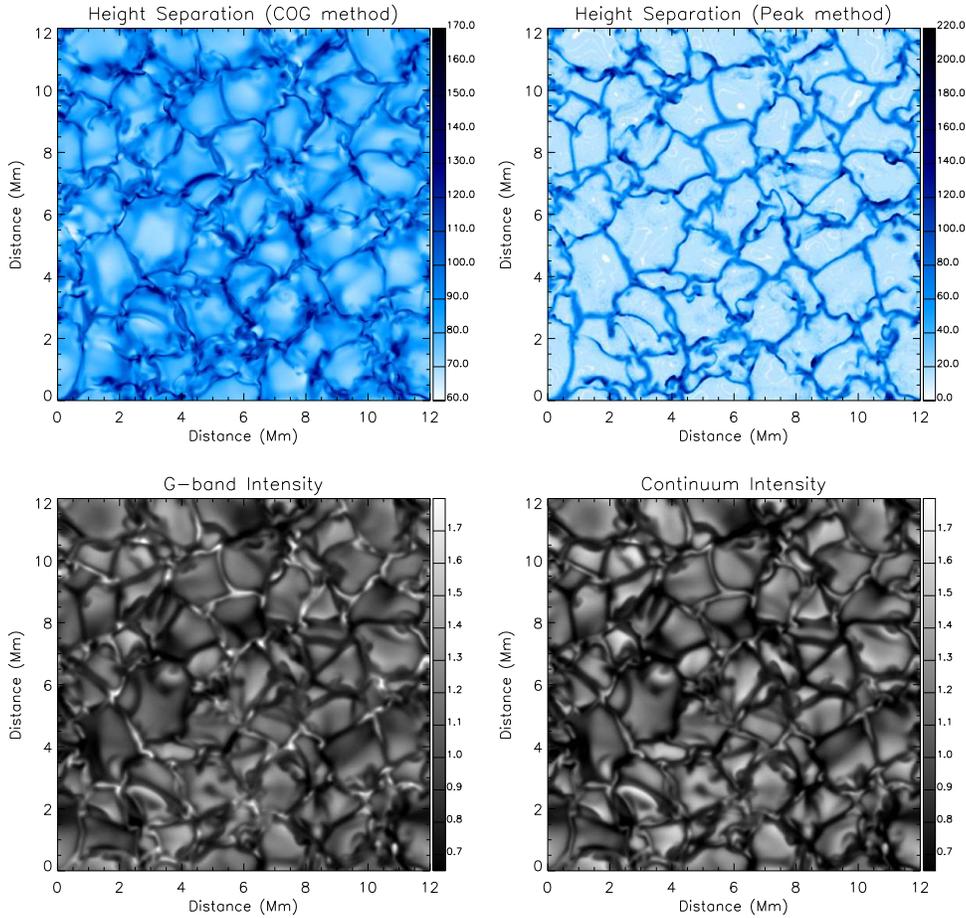}
\caption{The height separation between G-band and 4170{\,}{\AA} continuum 
bandpasses using the center-of-gravity (upper left) and peak subtraction 
(upper right) methods. In both upper panels, the color scale represents the 
separation distance (in km) using each of the methods detailed in 
\S~\ref{analysis}. Simultaneous G-band and 4170{\,}{\AA} continuum intensity 
images, normalised to their mean, are shown for reference in the lower 
left and right panels, respectively. All axes are distances along the solar surface, 
where 1~Mm $=$ 1000~km $\simeq$ 1.38$''$.
\label{fig6}}
\end{figure*}

The simulated G-band and 4170{\,}{\AA} continuum time series, first interpolated on to a 
constant-cadence grid, were subjected to the same rigorous wavelet and 
MBP detection routines as described above. Following the evaluation of 
period, power, coherence, and phase angle between the simulated data, 
we are able to create a simulation-specific occurrence plot as a function of period and 
phase angle, as for the observational data. The middle panels of Figure~\ref{fig4} 
show the occurrence distributions 
for oscillations generated in simulated MBP (left) and non-MBP (right) regions. As 
with the observational time series (upper panels of Fig.~\ref{fig4}), it is 
clear that a negative phase shift appears preferential. The peak occurrences are 
consistent with the observational data, at a phase angle of $-9{\degr}$. 
In the simulated occurrence plot, 96\% of MBP structures display a negative 
phase shift, compared with 95\% for non-MBP regions. Also 
of interest is the number of pixels undergoing oscillations in both the observational 
and simulated time series. Incorporating both MBP and non-MBP oscillations, a 
total of $6.06\times10^{7}$ observational pixels (0.8\% of the $7\times10^{9}$ total 
pixels) are observed to oscillate, compared with $1.62\times10^{7}$ simulated 
pixels (8.1\% of the $2\times10^{8}$ total pixels). The higher occurrence rate of 
oscillations in the simulated data is most likely a result of the noise-free time series. 
Detector dark current, readout noise, and processing abnormalities will reduce the 
coherence levels of oscillations detected in our observational dataset, 
and thus place them below our strict 0.85 threshold. Also noticeable in the middle 
panels of Figure~\ref{fig4} is the reduced spread of phase angles for the simulated 
occurrence plots. To investigate whether this is a direct consequence of the 
noise-free time series, noise consistent with the observational dataset 
was added, and the occurrence plots recreated. The acquired dataset has a 
combined photon and read noise of $\sim$1\% with respect to the observational 
count rates. The resulting image signal-to-noise ratio, $SNR$, can be calculated as 
$SNR = \mu / \sigma$, where $\mu$ is the signal mean and $\sigma$ is the 
standard deviation of the noise \citep[]{Sch00}. A signal mean of $\approx$5000 counts, 
combined with a dark frame standard deviation count of $\approx$19, provides an 
observational $SNR \approx 260$. The addition of 1\% noise to the simulated 
dataset did not modify the occurrence plots by a significant margin. However, when the 
noise level was increased to 5\%, the peak occurrence phase angle and oscillation 
period remained consistent with the noise-free time series, but with a significantly 
larger spread of phase angles (lower panels of Fig.~\ref{fig4}). 
A noise level of 5\% provides a simulated $SNR \approx 150$, still well above the 
Rose threshold for feature disambiguation \citep[]{Ros48}. Thus, simulated 
features remain well defined following the addition of 5\% noise, yet display a 
marked increase in the spread of recovered phase angles between the two filter 
bandpasses. Under these conditions, 80\% of 
MBP structures display a negative phase shift, compared with 79\% for non-MBP 
regions, indicating a closer resemblance to the observational time series. 
Again, these results imply that, even in the simulated domain, there is a 
preference for 4170{\,}{\AA} continuum oscillations to lead their G-band counterparts. 

\begin{figure*}
\epsscale{0.85}
\plotone{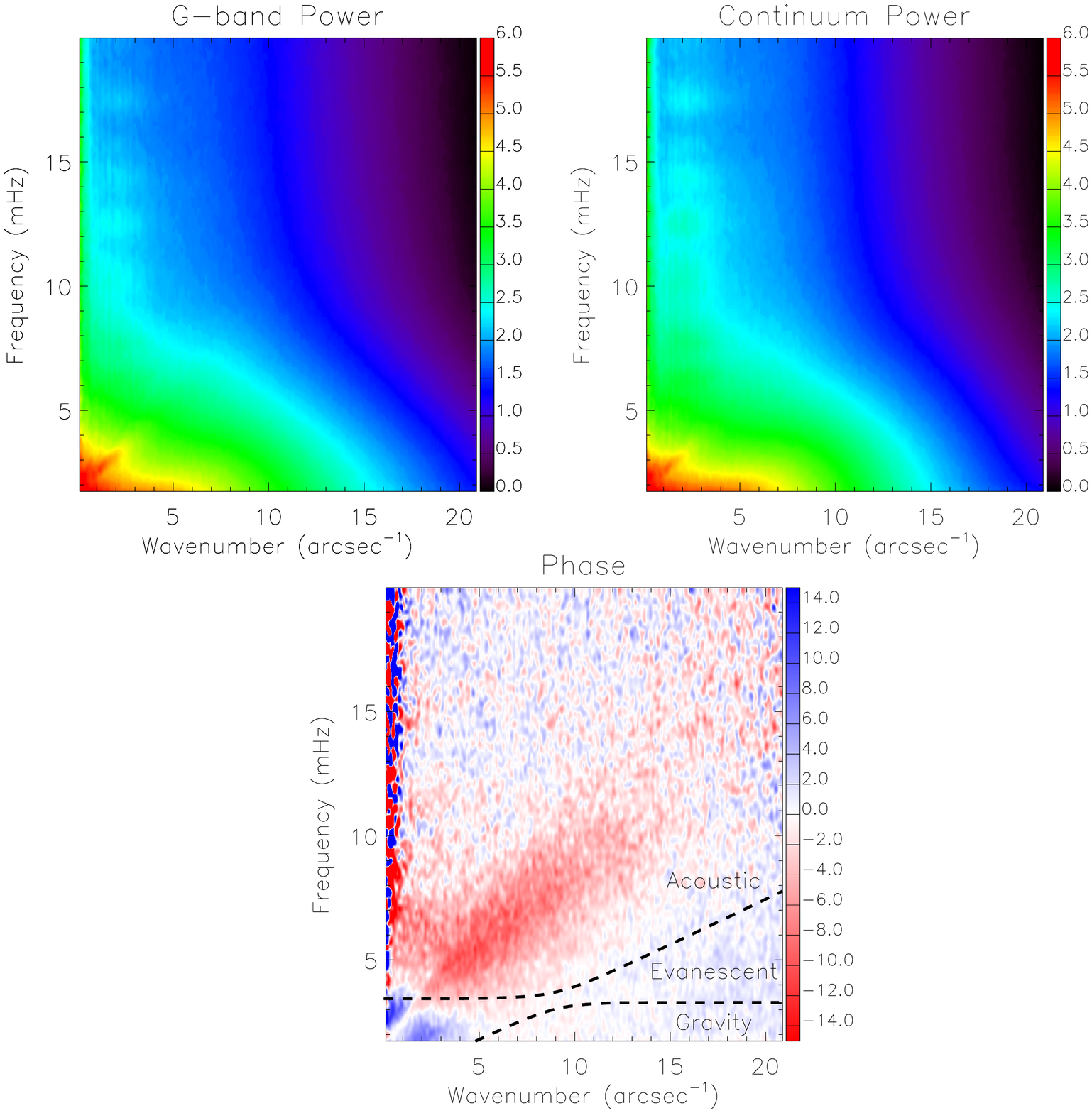}
\caption{Power spectra of the G-band (upper left) and 4170{\,}{\AA} continuum (upper right) 
observations as a function of frequency and spatial wavenumber. The color scale 
represents the power (in orders of magnitude) above the quiescent background. The lower 
panel shows the phase delay between the G-band and 4170{\,}{\AA} continuum oscillations 
as a function of frequency and spatial wavenumber. The color scale is in degrees, where 
positive and negative values indicate that the G-band leads, and trails, the 4170{\,}{\AA} 
continuum, respectively. Limiting curves for acoustic- and gravity-wave regimes are displayed 
as dashed lines, with an evanescent region sandwiched between the acoustic and gravity 
regimes. A similar figure, only for the simulated time series, is shown in 
Figure~\ref{fig8}.
\label{fig7}}
\end{figure*}

An important point to raise is how these 
oscillations are generated. Often, oscillations are produced in simulated time 
series through the addition of an input driver \citep[e.g.][]{Bog03, Cal08, Fed11b}. 
However, in this instance, no specific driver has been added to our time series. 
This implies that longitudinal oscillations readily arise as a natural consequence of 
convective motions, and their interaction with the magnetic field in the solar photosphere 
\citep[see e.g.][]{Kit11, She11a}. Such oscillations have been studied observationally by 
\citet{Goo98}, who suggested these signatures are the result of events occurring 
just beneath the photosphere, which subsequently convert into atmospheric 
longitudinal oscillations. Theoretical approaches \citep[e.g.][]{Ska00} have 
indicated that transient acoustic waves are excited at the top of the convective 
zone, and immediately above convective overshoot regions. Thus, 
the detection of an abundant array 
of frequencies over a range of photospheric structures, including MBPs and intergranular 
lanes, coupled with the rapid development of these oscillatory phenomena, further 
indicates how the ubiquity of oscillations may be traced back to magnetoconvective 
processes occurring throughout the solar photosphere. 

\subsubsection{Determination of formation heights}
The contribution functions for our G-band and 4170{\,}{\AA} continuum 
filters are displayed in Figure~\ref{fig5}, for both MBP and granular structures, in 
addition to an average of the entire field-of-view. Our numerical domain is positioned 
in such a way that the visible solar surface, roughly defined as the horizontal geometrical 
layer which is physically close to the optical layer of radiation formation, is 
located approximately 600~km below the upper boundary \citep[i.e. at a domain height 
of 800~km;][]{She11b}. Initially, 
a peak subtraction method was utilised to estimate the height separation between 
the two filter bandpasses. This process involved establishing the height at which the 
4170{\,}{\AA} continuum exhibited its largest contribution, and subtracting this from 
the simultaneous (and co-spatial) peak contribution height for the G-band filter. This 
always resulted in positive distances, ranging from approximately 0~km at the 
center of granules, to over 200~km in some intergranular lanes and MBPs (upper-right 
panel of Fig.~\ref{fig6}). However, from Figure~\ref{fig5}, it is clear that the contribution 
functions for both G-band and 4170{\,}{\AA} continuum filters are considerably 
asymmetric. A prolonged tail, present in the G-band contribution functions, suggests 
that more emphasis should be placed on these extended heights, in order to accurately 
represent potential height separations. Thus, a center-of-gravity (COG) method was 
chosen to provide a more robust calculation of the distance between the two 
filter bandpasses. For each bandpass, the COG was calculated from the 
corresponding contribution function (CF) using,
\begin{equation}
\mathrm{COG} = \frac{\sum_{i=0}^{z} \left(\mathrm{CF}_{i} \times i\right)}
{\sum_{i=0}^{z} \mathrm{CF}_{i}} \ ,
\end{equation}
where $i$ represents individual measurements along the simulation's vertical 
domain, up to a maximum height, $z$, corresponding to 1.4~Mm. The COG 
measurement for the 4170{\,}{\AA} continuum bandpass was then subtracted 
from the simultaneous (and co-spatial) COG value derived for the G-band filter. 
Under this regime, all estimated distances were again positive. However, now 
a minimum height separation of approximately 60~km exists, even for 
contribution functions acquired at the center of granular cells, where the 
previous method estimated a 0~km separation. This is a direct result of the 
inclusion of the extended G-band contribution tail seen in Figure~\ref{fig5}. 
The upper-left panel of Figure~\ref{fig6} displays a 2-dimensional representation of 
the estimated atmospheric height separation between G-band and 4170{\,}{\AA} 
continuum filter bandpasses using the COG method. Interestingly, this approach 
does not provide the same maximum separation distance as found in the peak 
subtraction method. Using the COG approach, a maximum separation of 
approximately 170~km was found for some intergranular lanes and MBP structures, 
compared to a distance of nearly 220~km using the peak separation method. 
Nevertheless, the two methods provide similar conclusions, whereby the 
G-band is formed above the 4170{\,}{\AA} continuum, particularly in regions 
where strong magnetic field concentrations are expected, such as 
intergranular lanes and MBPs. In these locations, an average height 
separation of $\approx$55~km and $\approx$83~km is found for the 
peak subtraction and COG methods, respectively. Furthermore, 
it is interesting to see how, in the locations of MBPs, the peaks of the 
G-band and 4170{\,}{\AA} continuum contribution functions lie below the 
layer corresponding to the visible solar surface. This is consistent with 
the detailed models of \citet{Spr76}, \citet{Sch03}, and \citet{shelyagbp2, She10}, who 
describe how the creation of MBPs is a result of the tenuous interplay 
between hot photospheric plasma and evacuated magnetic flux tubes.

\begin{figure*}
\epsscale{0.85}
\plotone{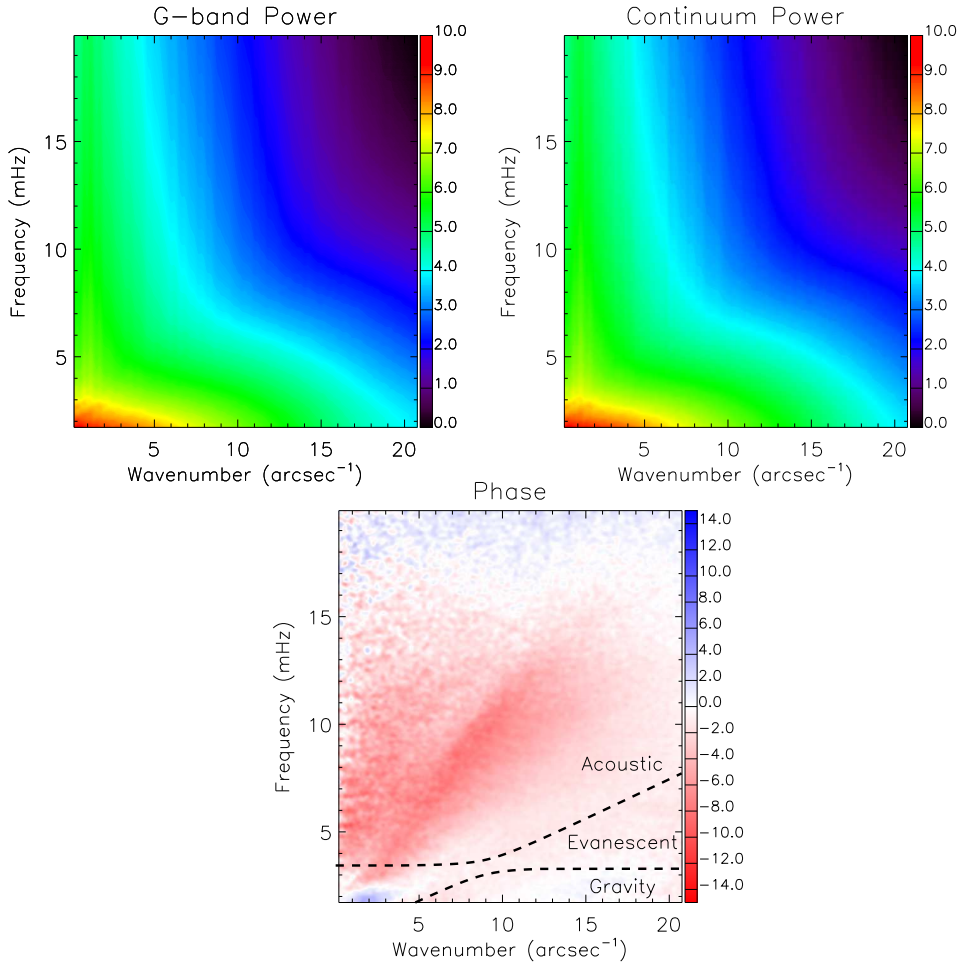}
\caption{Power spectra of the G-band (upper left) and 4170{\,}{\AA} continuum (upper right) 
simulations as a function of frequency and spatial wavenumber. The color scale 
represents the power (in orders of magnitude) above the quiescent background. The lower 
panel shows the phase delay between the G-band and 4170{\,}{\AA} continuum oscillations 
as a function of frequency and spatial wavenumber. The color scale is in degrees, where 
positive and negative values indicate that the G-band leads, and trails, the 4170{\,}{\AA} 
continuum, respectively. Limiting curves for acoustic- and gravity-wave regimes are displayed 
as dashed lines, with an evanescent region sandwiched between the acoustic and gravity 
regimes. A similar figure, only for the observational time series, is shown in 
Figure~\ref{fig7}.
\label{fig8}}
\end{figure*}

The formation height for each filter bandpass can also be estimated 
by applying these techniques to the averaged field-of-view contribution 
functions. Using the peak subtraction method, a formation height of 
$\sim$85~km and $\sim$40~km above the visible surface (located at $\sim$800~km 
in Fig.~\ref{fig5}) is found for 
G-band and 4170{\,}{\AA} continuum filters, respectively. When the COG 
method is utilised, the formation heights of G-band and 4170{\,}{\AA} continuum 
bandpasses becomes modified slightly to $\sim$100~km and $\sim$25~km, 
respectively. It must be noted that these are lower limits of the height of 
formation. Molecular and atomic line absorption profiles, which exist in the filter's 
spectral bandpasses, will contribute to the overall radiation absorption, thus shifting 
the total (continuum $+$ line) contribution functions. As a result, this will push 
radiation formation ranges upwards towards the cooler regions of the 
solar photosphere. Consequently, while the G-band will continue to be formed above the 
4170{\,}{\AA} continuum, the exact formation heights will be based on the 
lower limits established above. This has important consequences for the phase 
angles determined between simultaneous G-band and 4170{\,}{\AA} continuum 
oscillations. Since the height separation between the 4170{\,}{\AA} continuum and 
the G-band is always positive, a negative phase shift can be attributed to an 
upwardly propagating wave, such as those studied numerically by 
\citet{Wed03, Wed04}. Conversely, a positive phase shift will suggest the presence 
of a downwardly propagating wave, with a potential explanation incorporating aspects of 
wave reflection. However, the discrepancy between the amount of reflected waves in 
the observational ($\sim$28\% with positive phase angles) and simulated ($\sim$5\% with 
positive phase angles) time series may be related to the upper boundary conditions 
applied to the numerical simulations, in addition to noise aspects discussed above. 
The simulated upper boundary is at least partially 
reflective, but rigidly corresponds to a constant height in the photosphere. Contrarily, 
in the real Sun, this reflective layer can vary as a function of time and spatial position, 
potentially leading to a broader scatter of phase differences for the reflected wave.

\subsubsection{Oscillation frequency as a function of spatial size}
While it is clear from Figure~\ref{fig4} that a spread of oscillation frequencies 
is present in the data, it is important to investigate how individual 
photospheric structures affect the detected frequency. Following the procedures detailed 
in \citet{Kri01}, power spectra for the G-band and 4170{\,}{\AA} continuum observations 
and simulations are shown as a function of oscillation frequency and spatial wavenumber in 
the upper panels of Figures~\ref{fig7} \& \ref{fig8}, respectively. In each spectra, the 
majority of oscillatory power is associated with low frequencies and large spatial scales, 
consistent with solar p-mode oscillations \citep[]{Duv88}. These lower-left portions of the 
observational and simulated spectra reveal power approaching six and ten orders of 
magnitude above the quiescent background, respectively, highlighting the considerable 
power associated with these waves. Moving to smaller spatial scales (i.e. higher wavenumbers), 
it is clear that higher oscillation frequencies tend to be associated with these structures. 
This effect is particularly pronounced in the upper panels of Figure~\ref{fig7}, where a 
diagonal band of high oscillatory power is seen to stretch diagonally away from the 
lower-left corner of the power spectra. This can be verified in Figure~\ref{fig4}, 
where oscillations down to $\approx$110~s can be detected in small-scale 
MBP structures, compared to a $\approx$160~s lower limit on typically larger, non-MBP 
regions. 

\begin{figure*}
\epsscale{1.0}
\plotone{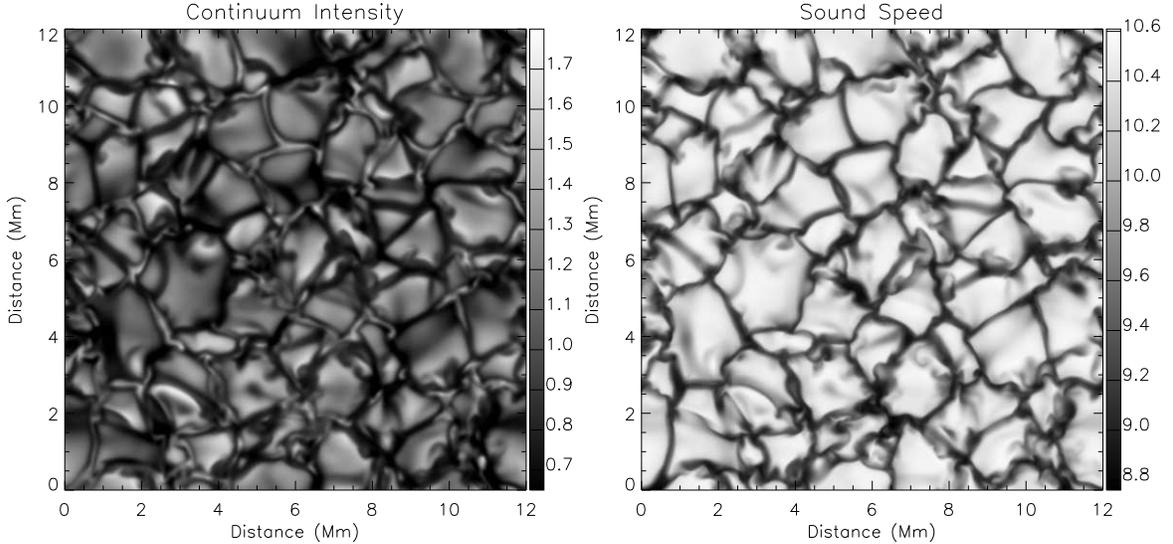}
\caption{A 4170{\,}{\AA} continuum intensity 
image (left), normalised to its mean, is displayed alongside the simultaneous 
pixelised sound speeds (right), where the color scale represents the derived sound speeds 
in km{\,}s$^{-1}$. The average sound speed for the entire simulated field-of-view is 10.1 km{\,}s$^{-1}$. 
All axes are distances along the solar surface, where 1~Mm $=$ 1000~km $\simeq$ 1.38$''$.
\label{fig9}}
\end{figure*}

It is interesting to note that almost no oscillations with periodicities below 
100~s are detected simultaneously in the G-band and 4170{\,}{\AA} continuum datasets. 
This applies to both the observational and simulated data, but may be related to the 
broad filter widths used. Recently, \citet{Kne11} 
detected upwardly propagating waves in the lower solar atmosphere with periodicities 
as short as 70~s. The authors utilised 2-D spectral imaging techniques to obtain 
a time series with an extremely narrow equivalent bandpass ($\approx$18~m{\AA}). 
Contrarily, the contribution functions extracted from our numerical 
simulations extend over a range of atmospheric heights due to the broad nature of 
our G-band and 4170{\,}{\AA} continuum filters, potentially decreasing 
and/or smoothing wave signals with periods shorter than 100~s. This does not 
necessarily mean that waves with periodicities below 100~s do 
not occur individually in either the G-band or 4170{\,}{\AA} continuum datasets. 
However, it does indicate that if these waves exist, we do not detect their propagation 
between atmospheric layers since no coherent signals can be determined.
The phase distribution displayed in the lower panel of 
Figure~\ref{fig7} reveals a shift of ${\approx}-$10${\degr}$ between 
G-band and 4170{\,}{\AA} continuum oscillations, apparent down to a spatial scale 
of $\sim$0.4$''$ (wavenumber 15~arcsec$^{-1}$). This spatial size is consistent 
with typical (roundish) MBP diameters, and forms 
an upper wavenumber limit beyond which the phase distribution becomes dominated 
by noise. These measurements are in agreement with the 
occurrence plots presented in Figure~\ref{fig4}, and reiterate the presence 
of upward propagating wave phenomena.

Examination of the lower panel of Figure~\ref{fig7} reveals a frequency 
``fork'' for wavenumbers below 4~arcsec$^{-1}$ (spatial size of approximately 
$1.6''$). This ``fork'' constitutes a continual linear downward trend of 
upwardly propagating waves (negative phases) for oscillation frequencies 
as a function of spatial wavenumber, in addition to a visible plateau at frequencies 
ranging from 4 -- 7~mHz, which includes the 3~min ($\approx$5.5~mHz) 
p-mode oscillation. A plateau suggests that the frequency of 
detected oscillations is no longer dependent on the increasing size of the 
spatial dimension, at least up to the $\approx$70$''$ limit imposed by our 
field-of-view. As a result, upwardly propagating 3~minute oscillations appear 
to be ubiquitous over granular (and larger) spatial scales. This agrees with 
the theoretical work of \citet{Erd07}, who demonstrate how 3~minute oscillations 
are free to propagate upwards from the solar surface as a consequence of 
their frequency residing above the acoustic cut-off value. A 3~minute periodicity overlaps with 
the solar p-mode spectrum, which has been shown by \citet{Mor01} to exist as a result 
of coherent pressure drivers up to spatial scales consistent with supergranular cells 
($\sim$50$''$). Thus, our results indicate an abundance of upwardly propagating 
longitudinal waves which cover the majority of our field-of-view, and suggests these 
wave forms are a direct result of a coherent (supergranular) p-mode driver. 
Interestingly, a similar phase plateau is not apparent in the simulated 
frequency-wavenumber diagram (lower panel of Fig.~\ref{fig8}). This may be a 
direct result of the reduced field-of-view of the simulated time series. With an overall 
size of $16.5 \times 16.5$~arcsec$^{2}$, compared to the observational field-of-view of 
$69.3 \times 69.1$~arcsec$^{2}$, information occurring on large spatial scales 
(i.e. low wavenumbers) will be lost. 

Evanescent waves are a phenomenon known 
to exist in the frequency-wavenumber diagram in locations of very low frequency 
and small wavenumber \citep[]{Fra68, Deu75a, Lit79, Deu90}. These waves cover the 
global 5 minute oscillation ($\sim$3~mHz), and are not expected to show signals 
of upward propagation. Positioned on either side of the evanescent region are 
locations in the frequency-wavenumber diagram where acoustic and gravity 
waves thrive \citep[]{Sou66, Sti70, Deu75b}. To investigate which regime 
best matches our detected wave phenomena, theoretical curves incorporating 
acoustic, evanescent, and gravity wave regimes, were over-plotted in the 
phase diagrams of Figures~\ref{fig7} \&~\ref{fig8}. In order to do this, the pressure, $P$, and 
density, $\rho$, were first extracted from our numerical simulations. Then, the 
sound speed, $c_s$, can be determined for each grid cell in our numerical domain using, 
\begin{equation}
c_{s} = \sqrt{\frac{\gamma P}{\rho}} \ ,
\end{equation}
%To do this, first the thermal energy was extracted 
%from our numerical simulations using the relation \citep[]{She08}, 
%\begin{equation}
%e_{\mathrm{thermal}} = e_{\mathrm{total}} - \frac{B^{2}}{2} - \frac{\rho v^{2}}{2} \ ,
%\end{equation}
%where $\rho$ is the density, $v$ is the velocity vector, and $B$ is the magnetic 
%field vector. Under this nomenclature, 
%$e_{\mathrm{thermal}}$, $e_{\mathrm{total}}$, $B^{2}/2$, and 
%$\rho v^{2}/2$ represent the thermal, total, magnetic, and kinetic 
%energy contributions, respectively. The thermal energy can then be related to the 
%gas pressure, $P$, via,
%\begin{equation}
%P = (\gamma - 1)~e_{\mathrm{thermal}} \ ,
%\end{equation}
where $\gamma$ is the ratio of specific heats, here taken as 5/3. 
%The sound 
%speed, $c_{s}$, can then be determined for each pixel in our numerical domain using,
%\begin{equation}
%c_{s} = \sqrt{\frac{\gamma P}{\rho}} \ .
%\end{equation}
The pixelised sound speeds were then convolved with our bandpass-specific 
contribution functions to determine an accurate representation of the sound 
speeds typically contained within our filtergrams. A two-dimensional representation 
of the derived sound speeds is shown in the right panel of Figure~\ref{fig9}, with 
an spatial average of 10.1~km{\,}s$^{-1}$. Using this value for the 
sound speed, and following the steps outlined in \citet{Kne11}, we 
derived the limiting curves for acoustic- and gravity-wave regimes. These 
frequency-wavenumber contours are over-plotted in the 
phase diagrams of Figures~\ref{fig7} \&~\ref{fig8}, with the locations of 
acoustic, evanescent, and gravity wave regimes annotated.
Close inspection of the lower panel in 
Figure~\ref{fig7} reveals how there are no upwardly propagating waves (i.e. 
negative phase angles) in areas of the diagram with frequencies less than 
$\sim$3~mHz (5~min) and wavenumbers less than $\sim$4~arcsec$^{-1}$ 
(1.6$''$), consistent with previous studies of evanescent waves. 
In this portion of the frequency-wavenumber diagram, only downwardly propagating 
waves appear to prevail. A similar trend is detected in the simulated 
time series (lower panel of Fig.~\ref{fig8}), where downwardly propagating 
waves dominate for wavenumbers less than $\sim$4~arcsec$^{-1}$ and 
frequencies less than $\sim$2.5~mHz. Slight differences between the 
observational and simulated frequency-wavenumber diagrams at these 
extreme limits of frequency and wavenumber may again be a result of the reduced 
field-of-view size and time series length for the simulated data.
Nevertheless, the observational and simulated 
phase distributions (lower panels of Figs.~\ref{fig7} \& \ref{fig8}) reveal identical 
trends, whereby propagating waves, best described as acoustic phenomena, 
with a similar phase angle increase in 
frequency with decreasing spatial size.

\subsubsection{Wave propagation velocity}
A phase difference of $-8{\degr}$ (as found in our observational time series), in 
conjunction with the most commonly occurring periodicity of 390~s, 
results in a time lag of 8.7~s between 
4170{\,}{\AA} continuum and G-band atmospheric heights. Utilising a traversed distance of 
55 -- 83~km, estimated through a comparison of G-band and 4170{\,}{\AA} continuum 
MBP contribution functions in our radiative MHD simulations, an average 
propagation speed of $7.9\pm1.6$~km{\,}s$^{-1}$ is found. This derived wave speed is 
close to typical photospheric sound speeds, and reveals how, at least in the 
photospheric regime, the detected oscillations are linear in nature.

\section{Concluding Remarks}

Here we present high-cadence observations and simulations of the solar 
photosphere. Each dataset demonstrates a wealth of oscillatory behaviour, 
with high concentrations found in highly magnetic regions, such as magnetic 
bright points (MBPs), visible as periodic intensity fluctuations with periods 
in the range 110 -- 600~s. Using a high-resolution radiative magneto-hydrodynamic 
(MHD) code to simulate photospheric magnetoconvection, we detect a wealth of 
upwardly propagating acoustic waves over a range of frequencies and wavenumbers. 
Results of the MHD simulations are consistent with those found in our observational 
time series, confirming that current MHD simulations are able to accurately 
replicate propagating wave phenomena in the lower solar atmosphere, 
without the need of an external driver. We suggest how the ubiquity of these 
oscillations may be traced back to magnetoconvective processes occurring in the 
upper layers of the Sun's convection zone. We are able to estimate the 
average height of formation of our G-band and 4170{\,}{\AA} continuum 
filters by convolving the filter bandpasses with the radiative output of our simulation. 
This provides minimum G-band and 4170{\,}{\AA} continuum formation heights of 
100~km and 25~km, respectively. We find that longitudinal oscillations 
exhibit a dominant phase delay of $-8{\degr}$ between G-band and 
4170{\,}{\AA} continuum observations, suggesting the presence of 
upwardly propagating waves. Almost no propagating waves with periods 
less than 100~s are detected in either observational or simulated data.
This does not necessarily mean that these high-frequency oscillations do 
not occur individually in either bandpass. However, it does indicate that if 
these waves exist, we do not detect their propagation between atmospheric layers 
since no coherent signals can be determined. More than 73\% of MBPs 
(73\% from observations, 96\% from simulations) display upwardly propagating 
wave phenomena, helping to explain the ubiquitous nature of MHD waves in the outer 
solar atmosphere.

\acknowledgments
DBJ thanks the Science and Technology Facilities Council (STFC) for a Post-Doctoral Fellowship. 
PHK is grateful to the Northern Ireland Department of Education and Learning for a PhD studentship. 
DJC thanks the CSUN College of Science for start-up funding related to this project. 
Solar Physics research at QUB is supported by STFC. 
The ROSA project is supported by The European Office of Aerospace Research \& Development.


%% in parentheses, after the keyword, but they will not be verified.

{\it Facilities:} \facility{Dunn (ROSA)}.

\clearpage

%% the file, uncomment the \plotone, \plottwo, and
%% If you need a layout that cannot be achieved with \plotone or
%% \plottwo, you can invoke the graphicx package directly with the
%% \includegraphics command or use \plotfiddle. For more information,
%% please see the tutorial on "Using Electronic Art with AASTeX" in the
%% documentation section at the AASTeX Web site, http://aastex.aas.org/
%%
%% The examples below also include sample markup for submission of
%% supplemental electronic materials. As always, be sure to check
%% the instructions to authors for the journal you are submitting to
%% for specific submissions guidelines as they vary from

%% This example uses \plotone to include an EPS file scaled to
%% 80% of its natural size with \epsscale. Its caption
%% has been written to indicate that additional figure parts will be
%% available in the electronic journal.

\clearpage

\clearpage

\clearpage

\clearpage

\clearpage

\clearpage

\clearpage

\clearpage

\end{document}